# Magnetism in the $KBaRE(BO_3)_2$ (RE=Sm, Eu, Gd, Tb, Dy, Ho, Er, Tm, Yb, Lu) series: materials with a triangular rare earth lattice

M. B. Sanders, F. A. Cevallos, R. J. Cava
*Department of Chemistry, Princeton University, Princeton, New Jersey 08544*

*Corresponding authors: marisas@princeton.edu (M.B. Sanders), rcava@princeton.edu (R.J. Cava)

**Abstract**

We report the magnetic properties of compounds in the $KBaRE(BO_3)_2$ family (RE= Sm, Eu, Gd, Tb, Dy, Ho, Er, Tm, Yb), materials with a planar triangular lattice composed of rare earth ions. The samples were analyzed by x-ray diffraction and crystallize in the space group *R-3m*. Physical property measurements indicate the compounds display predominantly antiferromagnetic interactions between spins without any signs of magnetic ordering above 1.8 K. The ideal 2D rare earth triangular layers in this structure type make it a potential model system for investigating magnetic frustration in rare-earth-based materials.

## Introduction

Geometrically frustrated magnetic materials are a fertile area of study in condensed matter physics. Competing interactions can lead to highly degenerate low temperature states and unconventional quantum phenomena, providing models for investigating exotic quasi-particle excitations and statistical mechanics. Built from corner-sharing magnetic tetrahedra, the 3D pyrochlore is a prime example of geometric frustration. These materials can exhibit a variety of magnetic properties, such as spin ice ($Dy_2Ti_2O_7$, $Ho_2Sn_2O_7$, $Ho_2Ti_2O_7$),[1,2] spin liquid ($Tb_2Ti_2O_7$),[3] long range ordered ($Gd_2Ti_2O_7$, $GdSn_2O_7$, $Er_2Ti_2O_7$),[4] and spin glass ($Tb_2Mo_2O_7$) behavior.[5]

Two-dimensional frustrated lattices are also of significant interest. In two dimensions, the simplest geometrically frustrated structure consists of a triangular lattice with a single magnetic ion per unit cell. This is the original system for which a 2D quantum spin liquid state was proposed over forty years ago.[6] Since then, extensive theoretical and experimental research has been dedicated to the search for novel 2D frustrated magnetic materials. Two new series of compounds containing ideal rare earth kagome planes have recently been reported, $RE_3Sb_3Mg_2O_{14}$ and $RE_3Sb_3Zn_2O_{14}$.[7,8] These materials display interesting magnetic ground states, including scalar chiral spin 1/2 order, kagome spin ice, dipolar spin order, and the KT transition.[9,10] Moreover, recent studies on $YbMgGaO_4$, which contains two-dimensional triangular layers of $Yb^{3+}$, suggest that the material is a potential spin liquid candidate.[11] The subtle interplay of the rare earth ion dependent crystal field anisotropy and frustrated lattice in these 2D materials can produce a variety of interesting magnetic phenomena. Realizing more rare earth-based 2D frustrated systems may lead to even more exotic physical properties.

Here we present the elementary magnetic properties and lattice parameters of $KBaRE(BO_3)_2$ (RE=Sm, Eu, Gd, Tb, Dy, Ho, Er, Tm, Yb, Lu), a series of materials with a 2D triangular lattice of rare earth ions. This structure type has previously been reported for RE= Tb, Y, Lu, Gd[12,13,14,15] The materials reported here are isostructural and crystallize with the Buetschliite [$K_2Ca(CO_3)_2$] mineral structure type in the space group *R-3m*. The distinct triangular layers constructed by $RE^{3+}$ in $KBaRE(BO_3)_2$ make this structure type a model system for exploring magnetic frustration on an ideal triangular lattice.

## Experimental

Samples were synthesized by solid state reaction using the rare earth oxides, $BaCO_3$, $K_2CO_3$, and $H_3BO_3$ as starting materials. The rare earth oxides ($Sm_2O_3$, $Eu_2O_3$, $Gd_2O_3$, $Tb_4O_7$, $Dy_2O_3$, $Ho_2O_3$, $Er_2O_3$, $Tm_2O_3$, $Yb_2O_3$, and $Lu_2O_3$) were dried at 800 °C overnight prior to use. The carbonates were dried at 120°C for several days. Stoichiometric mixtures (with 5% excess $H_3BO_3$ and 3% excess $K_2CO_3$ and $BaCO_3$) were ground thoroughly in an agate mortar and pestle, placed in alumina crucibles, and pre-reacted in air at 500 °C for 15 hours. Following this, the samples were reground, reacted in air at 900 °C for 20 hours, and then furnace-cooled to room temperature. $KBaTb(BO_3)_2$ was heated under flowing argon at 900 °C for an additional 10 hours. The X-ray powder diffraction data were collected at room temperature using a Bruker D8 Advance Eco diffractometer with Cu Kα radiation (λ=1.5418 Å) and a Lynxeye detector. Le Bail fits on $KBaRE(BO_3)_2$ were performed with the program Fullprof. The magnetic susceptibilities for the $KBaRE(BO_3)_2$ materials were measured between 1.8 and 300 K in a Quantum Design Physical Properties Measurement System (PPMS) Dynacool in an applied field of 5000 Oe. The magnetizations were linearly proportional to the magnetic field for all temperatures above 1.8 K to up to fields of approximately 15,000 Oe in all materials. The magnetic susceptibility was therefore defined as M/H at the field of H=5000 Oe.

**Results and Discussion**

*Crystal Structure of KBaRE(BO₃)₂*

The crystal structure of $KBaRE(BO_3)_2$ has been previously reported and is shown in Figures 1 and 2.[16,13,12,15] The materials reported here are all isostructural and crystallize in the space group *R-3m.* Several Le Bail profile fits to powder diffraction data from our samples are shown in Figures 3 and 4, while the lattice parameters are provided in Table 1. In the structure of $KBaRE(BO_3)_2$ (Figure 1), layers of flat $BO_3$ triangular units extend orthogonal to the crystallographic c axis. Interlinked between these layers are double sheets of Ba/K atoms and single sheets of rare earth atoms. The rare earth layers are fully structurally ordered. Each $3^+$ rare earth ion is octahedrally coordinated to oxygen, while the $Ba^{2+}$ and $K^+$ ions are mixed on the same site in a 9-coordinated Ba/K-O polyhedron. As emphasized in Figure 2, the rare earth ions make up discrete triangular layers in the structure. The lattice parameters of the title compounds agree well with those previously reported.[13,14,12,15] As expected, the unit cell parameters decrease with decreasing ionic radius of the rare earth, as shown in Figure 5.[17]

*Magnetic Properties*

The magnetic data for $KBaRE(BO_3)_2$ was fit to the Curie-Weiss law $\chi = \frac{C}{T-\theta_{CW}}$ where $\chi$ is the magnetic susceptibility, C is the Curie Constant, and $\theta_{CW}$ is the Weiss temperature. The effective moments were then obtained using the following relationship: $\mu_{eff} \propto 2.83\sqrt{C}$. The parameters for the fits are presented in Table 2, along with comparisons to the expected moments for each $RE^{3+}$ ion.[18] At high temperatures, crystal field effects may influence magnetic correlations and thereby impact susceptibility measurements; therefore, data was fit to the Curie-Weiss law at both low and high temperatures for comparison. Figures 6-14 feature Curie-Weiss fits of the samples reported here, along with field-dependent magnetization curves for the compounds at 2 K.

**$KBaSm(BO_3)_2$**

The high temperature susceptibility of $KBaSm(BO_3)_2$ in Figure 6 reveals a large temperature independent Van Vleck paramagnetism contribution. This is characteristic of many Sm-containing compounds and arises from the first excited J=7/2 multiplet of $Sm^{3+}$.[19] Therefore, the Curie-Weiss law was fit at the low temperature regime between 1.8 and 10 K, yielding a moment of 0.44 $\mu_B$/Sm and a Weiss temperature of -1.35 K. The negative Weiss temperature indicates the presence of antiferromagnetic nearest neighbor spin–spin interactions. While the $\mu_{eff}$ is somewhat smaller than the expected free ion value of 0.83 $\mu_B$/Sm, it is similar to that of other reported $Sm^{3+}$-containing oxides, such as $Sm_3Sb_3Mg_2O_{14}$ (0.53 $\mu_B$/Sm) and $Sm_2Zr_2O_7$ (0.50 $\mu_B$/Sm).[8,20] An effective moment as low as 0.15 $\mu_B$ has been reported for $Sm_2Ti_2O_7$.[20] The low moments observed for these compounds are likely related to samarium's large crystal field splitting of its lowest J=5/2 multiplet. The field-dependent magnetization *M(H)* at 2 K in the left panel of Figure 6 shows slight curvature but no indication of saturation up to an applied field of $\mu_0H$= 9 T.

**$KBaEu(BO_3)_2$**

For $Eu^{3+}$-containing compounds, typically $\mu_{eff}$ = 0. This arises because $\mu_{eff} = g_J[J(J + 1)]$, where for $Eu^{3+}$, L = 3 and S = 3 and so J = 0. It is expected, therefore, that the ground state for $Eu^{3+}$ may be non-magnetic, $^7F_0$. At low temperatures (up to roughly ~100 K), $KBaEu(BO_3)_2$'s susceptibility displays Van Vleck paramagnetism (Figure 7). This is characteristic of $Eu^{3+}$ compounds, and is a result of sole population of the non-magnetic ground state. As the temperature rises, crystal field states originating from the first excited multiplet, $^7F_1$, are populated. This ultimately leads to a temperature-dependent contribution to $KBaEu(BO_3)_2$'s magnetic susceptibility. Although this effect may appear Curie-Weiss-like, a fit of the data does not accurately reflect the exchange interactions, and so we do not apply the Curie-Weiss law to $KBaEu(BO_3)_2$. The field-dependent magnetization for $KBaEu(BO_3)_2$ at 2 K is shown in the right panel of Figure 7. The *M(H)* plot is linear and reversible with the field increasing and decreasing.

**$KBaGd(BO_3)_2$**

The temperature dependent magnetic susceptibility of $KBaGd(BO_3)_2$ is shown in Figure 8. As can be seen by both the susceptibility plot and the tabulated data in Table 2, this material maintains strict Curie-Weiss law behavior to the lowest temperatures studied, with Weiss temperatures relatively small, on the order of 1 K. Indeed, fitting the high temperature data (150-300 K) yields a magnetic moment of 7.70 $\mu_{B/}$Gd and a Weiss temperature of 1.64 K. A low temperature fit (1.8-25 K) provides the same effective moment, but a negative Weiss temperature of -0.78 K. The moment calculated here is close to the expected value of 7.94 $\mu_{B/}$Gd for the free ion $^8S_{7/2}$ ground state of $Gd^{3+}$. Similar moments with Weiss temperatures of a comparable magnitude were observed for the rare earth double perovskites $Ba_2GdSbO_6$ ($\mu_{eff}$=7.72; $\Theta_w$=-1.22 K) and $Sr_2GdSbO_6$ ($\mu_{eff}$=7.64; $\Theta_w$=-0.14 K).[21] The *M(H)* plot at 2 K in the right panel of Figure 8 reveals a nonlinear relationship between the magnetization and applied field and saturation at roughly 7.00 $\mu_{B/}$Gd, slightly less than the moment calculated through the Curie-Weiss law and representative of all the Gd spins aligning in the powder sample in the direction of the applied field by roughly $\mu_0H$= 5 T.

**$KBaTb(BO_3)_2$**

The $\chi(T)$ plot of $KBaTb(BO_3)_2$ is shown in Figure 9. A high temperature fit of the data (150-300 K) gives an effective moment of 9.65 $\mu_B$/Tb and a Weiss temperature of -15.06 K. A similar $\mu_{eff}$ of 9.69 and $\Theta_w$ of -11.09 were extrapolated from the low temperature regime. The moments calculated here are close to the expected free ion value of 9.72 $\mu_B$/Tb. The antiferromagnetic Curie-Weiss temperature obtained from the high temperature fit is close to that observed for the pyrochlore $Tb_2Ti_2O_7$ ($\Theta_w$= ~17 K).[22] The 2 K field dependent magnetization plot of $KBaTb(BO_3)_2$ displayed in the right panel of Figure 9 reveals a nonlinear response with some curvature but no signs of saturation up to an applied field of $\mu_0H$= 9 T.

**$KBaDy(BO_3)_2$**

Shown in Figure 10 is the magnetic susceptibility and inverse susceptibility of $KBaDy(BO_3)_2$. A high temperature (150-300 K) Curie-Weiss fit of the data yields a moment of 10.63 $\mu_B$/Dy and a Weiss temperature of -16.12 K, in agreement with the free ion moment of 10.63 $\mu_B$ expected for $Dy^{3+}$. The susceptibilities deviate toward smaller values at the lowest temperatures measured, suggesting increasing antiferromagnetic correlations and perhaps emergent long-range order.[21] Indeed, fitting at low temperatures provide $\mu_{eff}$= 10.07 $\mu_B$/Dy and $\Theta_w$=-5.42. This low temperature moment is close to the value of $\mu_{eff}$= 10.00 $\mu_B$, for a ground state mJ = ±15/2 doublet, with gJ = 4/3 and corresponds well with that of $Dy_2Ti_2O_7$.[23] The negative Weiss temperatures indicate antiferromagnetic interactions between spins. The field dependent magnetization at 2 K (Figure 10, right panel) reveals nonlinear variation of the magnetization as a function of applied field and the onset of saturation at roughly 5 $\mu_B$/Dy, about half the expected maximum of the effective moment. Saturation at about half the value of the effective moment is typically due to powder averaging of Ising spins.

**$KBaHo(BO_3)_2$**

Displayed in Figure 11 is the $\chi(T)$ plot of $KBaHo(BO_3)_2$. The susceptibility behavior of the Dy analog is similar to that of $KBaHo(BO_3)_2$. A high temperature (150-300 K) fit of the inverse susceptibility results in a magnetic moment of 10.50 $\mu_B$/Ho and a Weiss temperature of -15.05 K. Applying the Curie-Weiss law to the low temperature regime yields the following parameters: $\mu_{eff}$= 10.09$\mu_B$/Ho and $\Theta_w$=-3.10 K. The effective moments correspond well with the expected free ion value of 10.60 $\mu_B$/Ho and the value expected for an mJ = ±8 doublet ground state, $\mu_{eff}$ = 10.00 $\mu_B$.[23] The field dependent magnetization *M(H)* plot at 2 K in the right panel of Figure 11 shows curvature with the onset of saturation at about 5.5 $\mu_B$/Ho.

**$KBaEr(BO_3)_2$**

The magnetic susceptibility of $KBaEr(BO_3)_2$ is plotted in Figure 12. Fitting the data from 150 to 300 K presents an effective moment of 9.57 $\mu_B$/Er and a Weiss temperature of -12.67 K, consistent with the free ion moment of $\mu_{eff}$= 9.59 $\mu_B$ for $Er^{3+}$ ($^4I_{15/2}$). A low temperature fit of the data yields $\mu_{eff}$=8.86 $\mu_B$ and $\Theta_w$=-3.09 K, in agreement with that of $\mu_{eff}$ = 9.0 $\mu_B$ for an mJ = ±15/2 doublet with gJ = 1.2.[23] The field dependent magnetization at 2 K in the right panel of Figure 12 shows a nonlinear response, approaching saturation at roughly 5.5 $\mu_B$/Er, a bit more than half the free ion value.

**$KBaTm(BO_3)_2$**

Exhibited in Figure 13 is the magnetic susceptibility for $KBaTm(BO_3)_2$. Data were fit to the Curie-Weiss law in the temperature range 200-300 K, yielding an effective moment of 7.58 $\mu_B$ and $\Theta_w$ of -29.37 K. Fitting the data to a lower temperature range (65-85 K) provides $\mu_{eff}$= 7.80 $\mu_B$ and $\Theta_w$= -34.61. The moments correspond well with that expected for $Tm^{3+}$ (7.57 $\mu_B$) and the negative Weiss temperatures are suggestive of antiferromagnetic interactions between spins.

Due to crystal-field level splitting, $Tm^{3+}$ ($^4f_{12}$) often takes on a singlet ground state. In a low symmetry site like that of $Tm^{3+}$ in $KBaTm(BO_3)_2$, *-3m*, the $^3H_6$ free ion state splits into a collection of single-ion crystal-field ground states where the strength and symmetry of the crystal field influence the energy separations.[24] As a result, the susceptibility features both a zero first order Zeeman contribution at low temperatures and a Van Vleck contribution due to the excited states. The anomaly in the $\chi(T)$ plot for $KBaTm(BO_3)_2$ at around 15 K may therefore be interpreted as a Van Vleck component to the magnetic susceptibility. This anomaly was observed in all preparations of the Tm variant.

The *M(H)* plot at 2 K in Figure 13 of $KBaTm(BO_3)_2$ reveals slight curvature without any signs of saturation up to an applied field of $\mu_0H$= 9 T.

**$KBaYb(BO_3)_2$**

Plotted in Figure 14 is the magnetic susceptibility of $KBaYb(BO_3)_2$. A high temperature fit of the data from 125-275 K gives an effective moment of 4.76 $\mu_B$/Yb and a Weiss temperature of -110.60 K. Fitting the data to the Curie-Weiss law in the lower 1.8-25 K temperature regime yields $\mu_{eff}$= 2.67 $\mu_B$ and $\Theta_w$= -0.84 K. Similar temperature dependent behavior has been observed for $Yb_2Ti_2O_7$.[23,25] This trend is consistent with the free ion moment $\mu_{eff}$= 4.54 $\mu_B$ of $Yb^{+3}$ at room temperature and the movement of the ions into a Kramers doublet ground state at low temperature, with reduced moment. This behavior at low temperature is indicative of planar spin anisotropy. The field-dependent magnetization at 2 K in Figure 14 sheds more light on this anisotropy and reveals a nonlinear magnetization response with the appearance of saturation at about 1.5 $\mu_B$/Yb.

## Conclusion

We have synthesized and characterized several compounds in the $KBaRE(BO_3)_2$ family (RE= Sm, Eu, Gd, Tb, Dy, Ho, Er, Tm, Yb, Lu). Le Bail profile fits were performed on the powder X-ray diffraction data from the compounds, from which the lattice parameters were extracted. They agree well with parameters previously reported in this structure type and show good evidence for the lanthanide contraction. Elementary magnetic property measurements show no signs of magnetic ordering above 1.8 K. Further low temperature characterization of these triangular rare earth materials, especially in single crystal form, is of significant interest; because many of these materials have been grown as single crystals[12-15], such studies are expected to be reasonably performed.

## Acknowledgments

This research was performed under the auspices of the Institute for Quantum Matter, fully supported by the US DOE BES grant DE-FG02-08ER46544.

## Tables

### Table 1. Lattice Parameters of $KBaRE(BO_3)_2$ determined from fits to room temperature powder X-ray diffraction data

| RE | RE Ionic Radius (Å) | a (Å) | c (Å) |
|---|---|---|---|
| **Sm** | 0.958 | 5.49112(4) | 18.1138(7) |
| **Eu** | 0.947 | 5.48299(9) | 18.0410(4) |
| **Gd** | 0.938 | 5.47536(4) | 17.9790(3) |
| **Tb** | 0.923 | 5.46342(6) | 17.8893(3) |
| **Dy** | 0.912 | 5.45536(2) | 17.8340(8) |
| **Ho** | 0.901 | 5.44827(3) | 17.7647(2) |
| **Er** | 0.890 | 5.43635(6) | 17.7018(4) |
| **Tm** | 0.880 | 5.42881(2) | 17.6358(8) |
| **Yb** | 0.868 | 5.41952(5) | 17.6029(4) |
| **Lu** | 0.861 | 5.41331(5) | 17.5726(3) |

### Table 2. Effective moments (*p*), Weiss temperatures ($\theta w$), and goodness of fit ($R^2$) determined by least-squares fitting of the Curie-Weiss law to the magnetic susceptibility data in figures 6-14

| RE | High T Fit | *p* (High T) | $\Theta w$ (High T) | $R^2$ of fit | Low T Fit | *p* (Low T) | $\Theta w$ (Low T) | $R^2$ of fit | *p* exp |
|---|---|---|---|---|---|---|---|---|---|
| Sm | - | - | - | - | 1.8-10 K | 0.43(9) | -1.35(3) | 0.99175 | 0.85 |
| Eu | - | - | - | - | - | - | - | - | 0.0 |
| Gd | 150-300 | 7.70(1) | 1.64(7) | 0.99994 | 1.8-25 K | 7.70(1) | -0.78(4) | 0.99994 | 7.94 |
| Tb | 150-300 | 9.65(1) | -15.06(2) | 0.99977 | 10-25 K | 9.68(9) | -11.09(7) | 0.99887 | 9.72 |
| Dy | 150-300 | 10.62(5) | -16.11(5) | 0.99971 | 15-30 K | 10.07(4) | -5.41(7) | 0.99881 | 10.63 |
| Ho | 150-300 | 10.50(1) | -15.05(7) | 0.99981 | 10-25 K | 10.09(1) | -3.09(9) | 0.99983 | 10.60 |
| Er | 150-300 | 9.57(2) | -12.66(7) | 0.99975 | 8-23 K | 8.86(5) | -3.09(3) | 0.99963 | 9.59 |
| Tm | 200-300 | 7.58(3) | -29.37(1) | 0.99915 | 65-85 K | 7.80(5) | -34.61(2) | 0.99939 | 7.57 |
| Yb | 125-275 | 4.76(4) | -110.59(9) | 0.99999 | 1.8-25 K | 2.67(5) | -0.83(7) | 0.99817 | 4.53 |
| Lu | - | - | - | - | - | - | - | - | 0.0 |

**Figure Captions**

**Figure 1.** The crystal structure of $KBaRE(BO_3)_2$, showing the coordination polyhedra of the rare earth ions and boron. The magenta polyhedra represent $REO_6$, while the triangular $BO_3$ units are indicated in orange. Oxygens are excluded for clarity, but are found at the vertices of the coordination polyhedra.

**Figure 2.** Schematic of $KBaRE(BO_3)_2$ structure displaying the positions of the metal ions in the unit cell. To the right of the schematic is an extended lattice showing the discrete $RE^{3+}$ triangular planes in the structure.

**Figure 3.** Le Bail profile fits for $KBaEu(BO_3)_2$ (left) and $KBaDy(BO_3)_2$ (right). The experimental pattern is in red, the calculated pattern in black, and the difference plot in blue. The green marks indicate Bragg reflections.

**Figure 4.** Le Bail profile fits for $KBaHo(BO_3)_2$ (left) and $KBaTm(BO_3)_2$ (right). The experimental pattern is in red, the calculated pattern in black, and the difference plot in blue. The green marks indicate Bragg reflections.

**Figure 5.** The *a* and *c* lattice parameters for $KBaRE(BO_3)_2$ as a function of rare earth ionic radius. The pink and black points represent the *a* and *c* parameters, respectively. The standard deviations are smaller than the plotted points and so error bars are excluded from the figure.

**Figure 6.** Left Panel: Magnetic susceptibility and inverse susceptibility of $KBaSm(BO_3)_2$ measured in an applied field of 5000 Oe. The Curie-Weiss fit is shown in black. The inset is a magnified view of the low temperature inverse susceptibility. The right panel displays the field-dependent magnetization at 2 K.

**Figure 7.** In the left panel is the temperature-dependent magnetic susceptibility of $KBaEu(BO_3)_2$ in an applied field of 5000 Oe. The inset displays the low temperature inverse susceptibility. The right panel exhibits the *M(H)* of the compound at 2 K.

**Figure 8.** The DC magnetic susceptibility and reciprocal susceptibility of $KBaGd(BO_3)_2$ measured in an applied field of 5000 Oe (left panel). The Curie–Weiss fits are shown in yellow. The plot in the right panel shows the field-dependent magnetization at 2 K. The inset reveals a magnified view of the compound's low temperature inverse susceptibility.

**Figure 9.** Temperature-dependent magnetic susceptibility of $KBaTb(BO_3)_2$ measured in an applied field of 5000 Oe (left panel). The Curie–Weiss fits are shown in black. The right panel displays the field-dependent magnetization at 2 K. The inset shows the low temperature inverse susceptibility.

**Figure 10.** The DC magnetic susceptibility and inverse susceptibility of $KBaDy(BO_3)_2$ measured in an applied field of 5000 Oe. The Curie–Weiss fits are shown in yellow. The plot to the right of the *M(T)* displays the magnetization as a function of applied field *M(H)* at 2 K. The inset reveals the low temperature reciprocal susceptibility.

**Figure 11.** Right panel: the magnetic susceptibility and inverse susceptibility of $KBaHo(BO_3)_2$ in an applied field of 5000 Oe. The Curie-Weiss fits are shown in black. Left panel: the field-dependent magnetization at 2 K. The inset shows a magnified view of the low temperature reciprocal susceptibility.

**Figure 12.** Temperature-dependent magnetic susceptibility of $KBaEr(BO_3)_2$ in an applied field of 5000 Oe (left panel). The Curie–Weiss fits are shown in black. To the right of the susceptibility is the field-dependent magnetization at 2 K. The inset displays the low temperature inverse susceptibility.

**Figure 13.** The DC magnetic susceptibility and inverse susceptibility of $KBaTm(BO_3)_2$ in an applied field of 5000 Oe (left panel). The Curie–Weiss fits are shown in green. Right panel: the field-dependent magnetization at 2 K. The inset reveals a magnified view of the low temperature reciprocal susceptibility.

**Figure 14.** Right panel: the magnetic susceptibility and reciprocal susceptibility of $KBaYb(BO_3)_2$ in an applied field of 5000 Oe. The Curie-Weiss fits are shown in black. Displayed in the left panel is the field-dependent magnetization at 2 K. The inset reveals the low temperature inverse susceptibility.

**Figure 1**

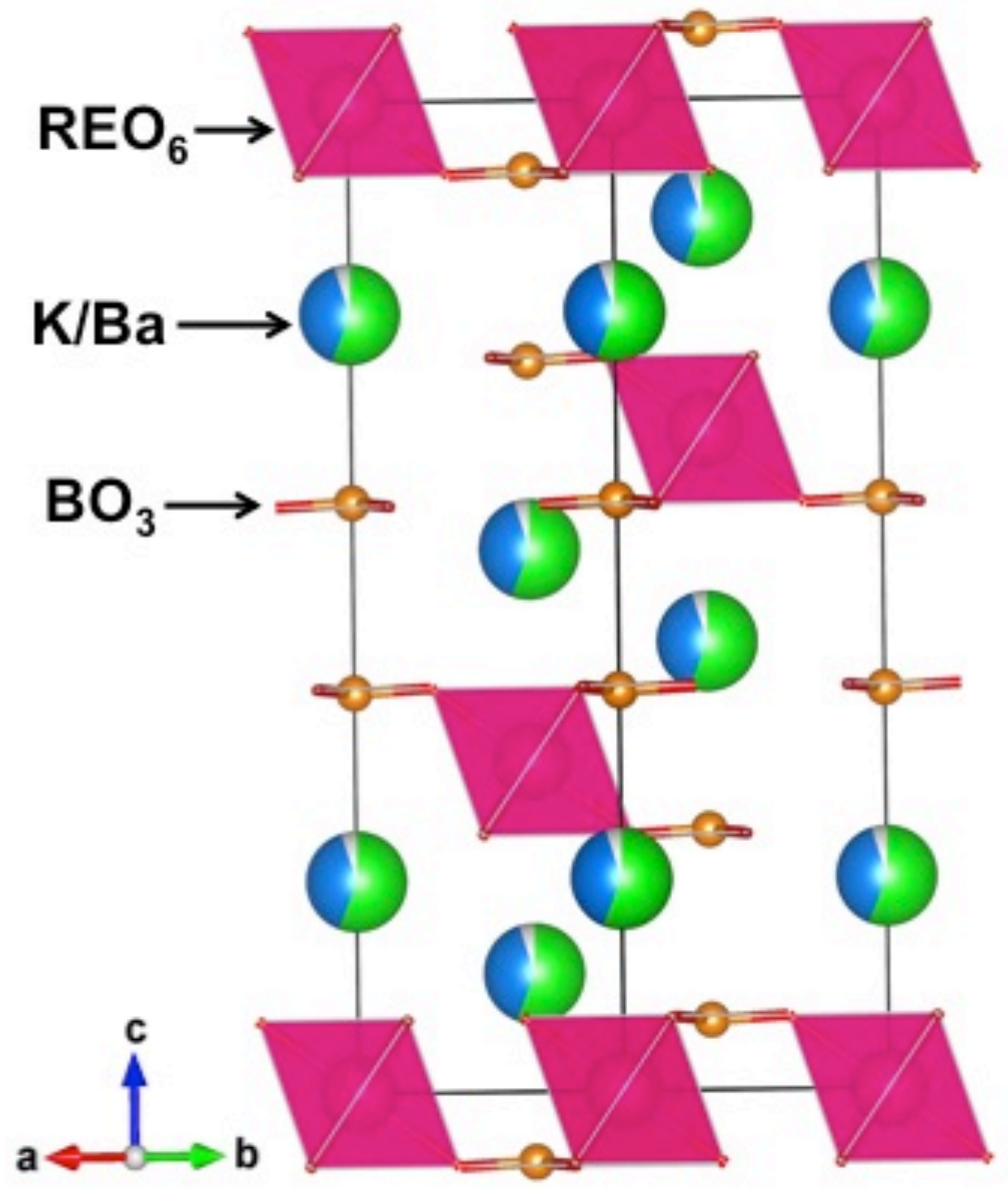

**Figure 2**

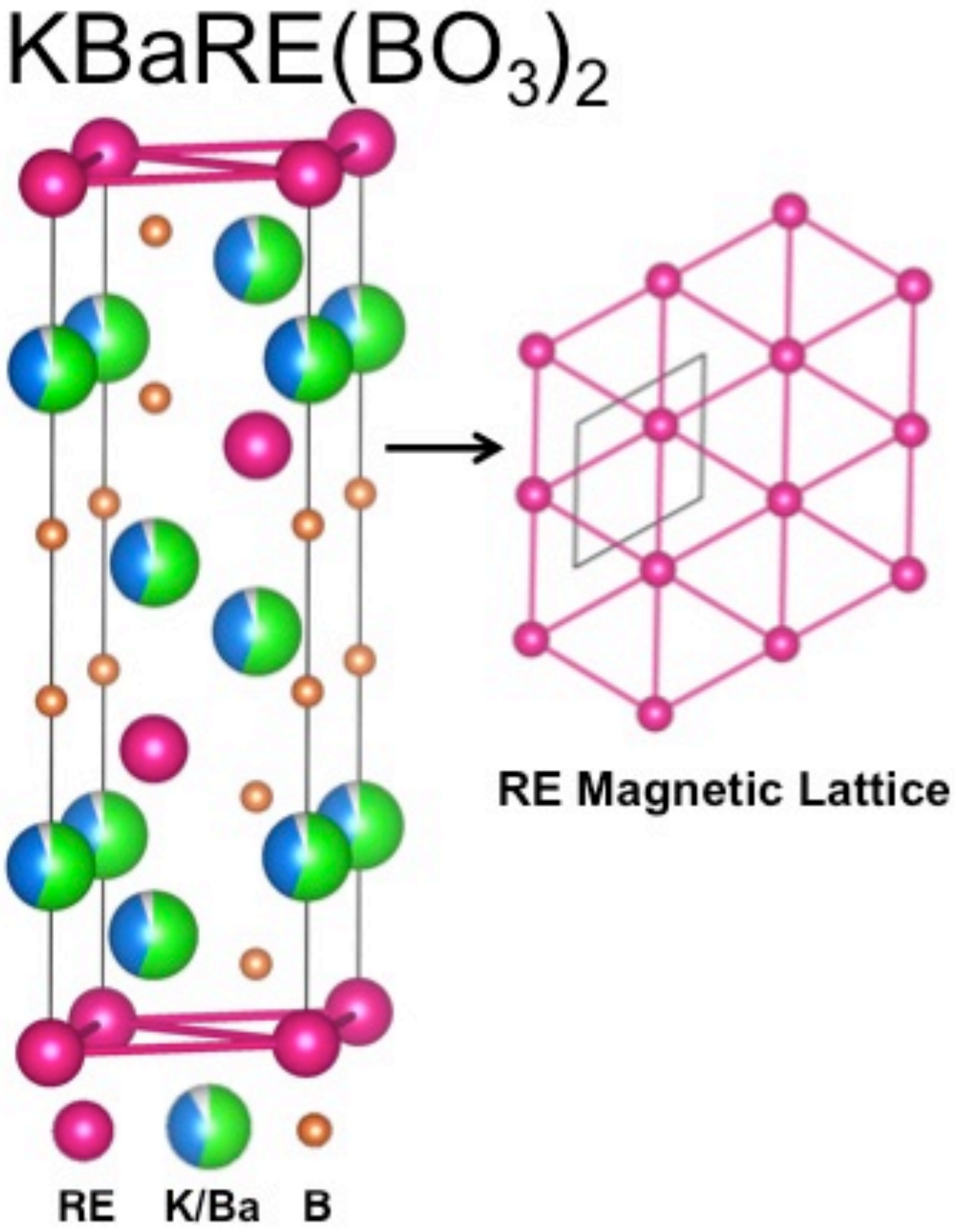

**Figure 3**

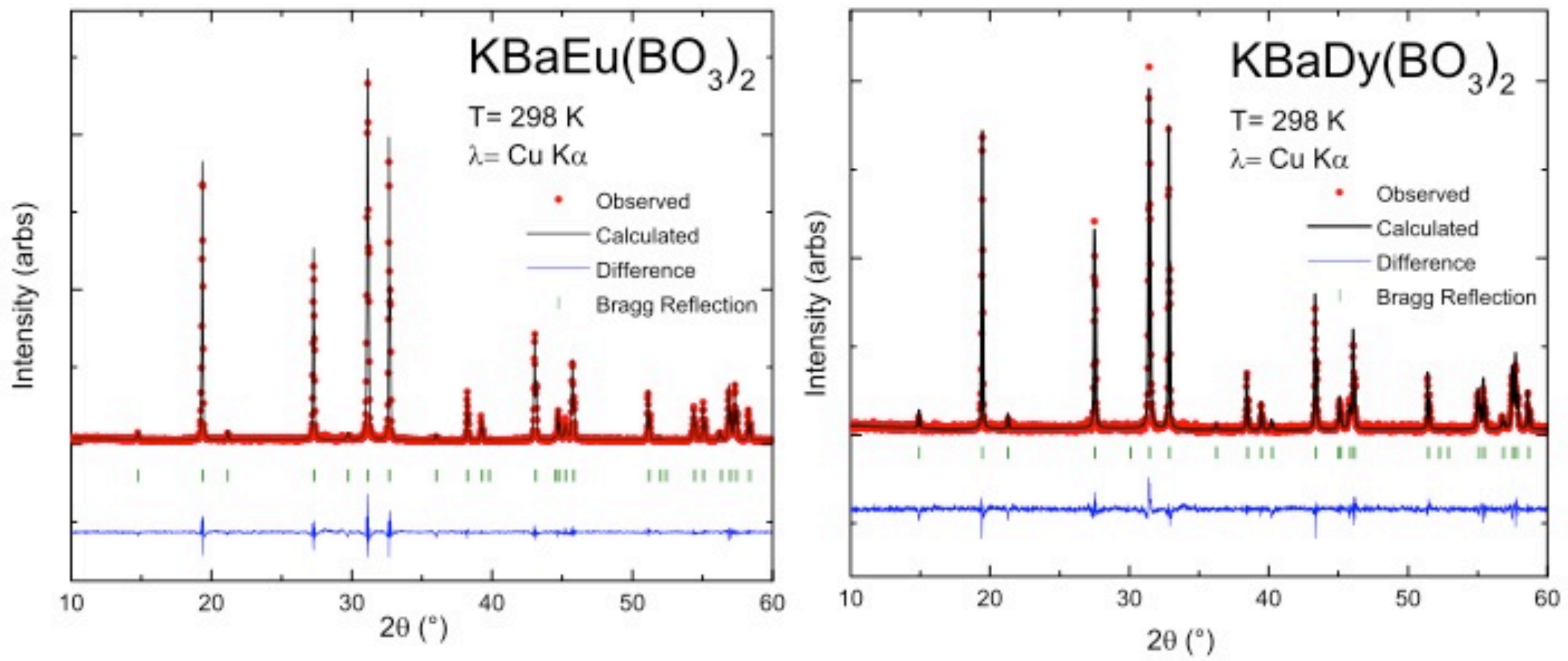

**Figure 4**

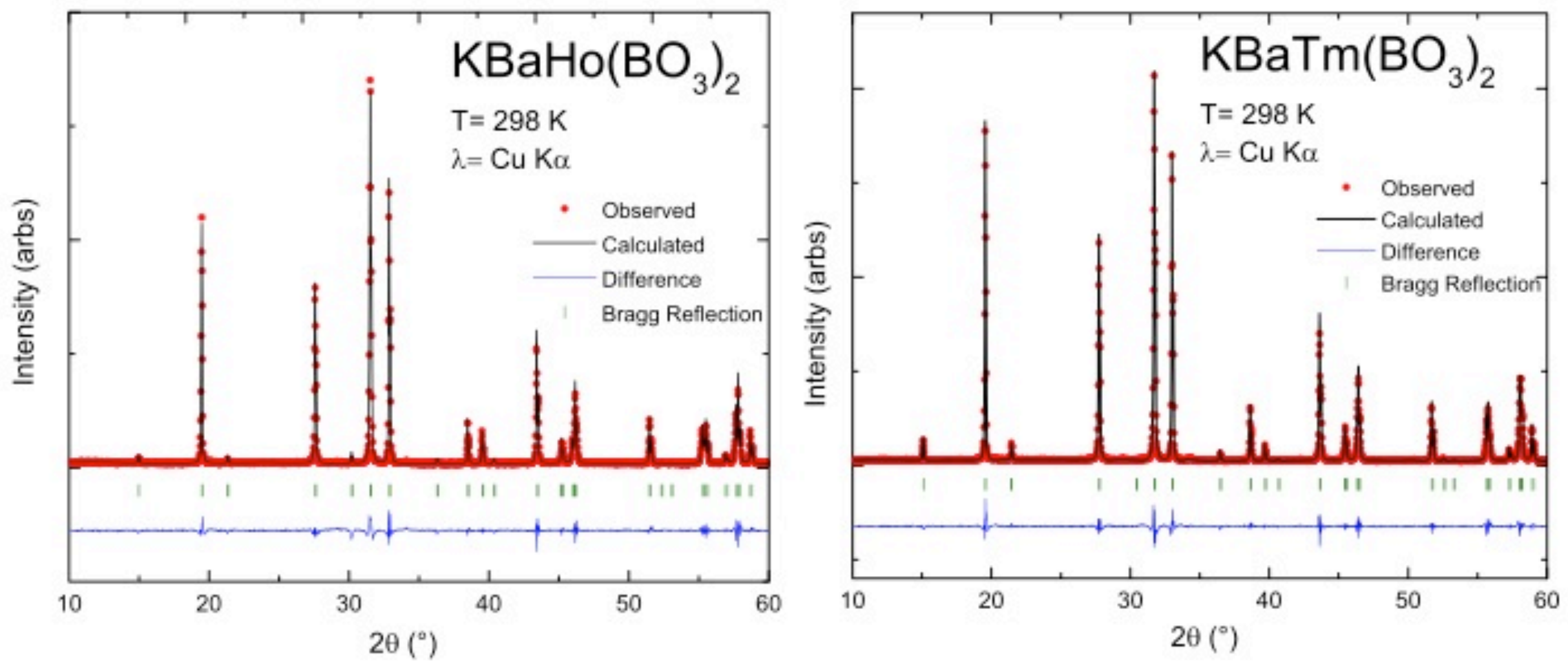

**Figure 5**

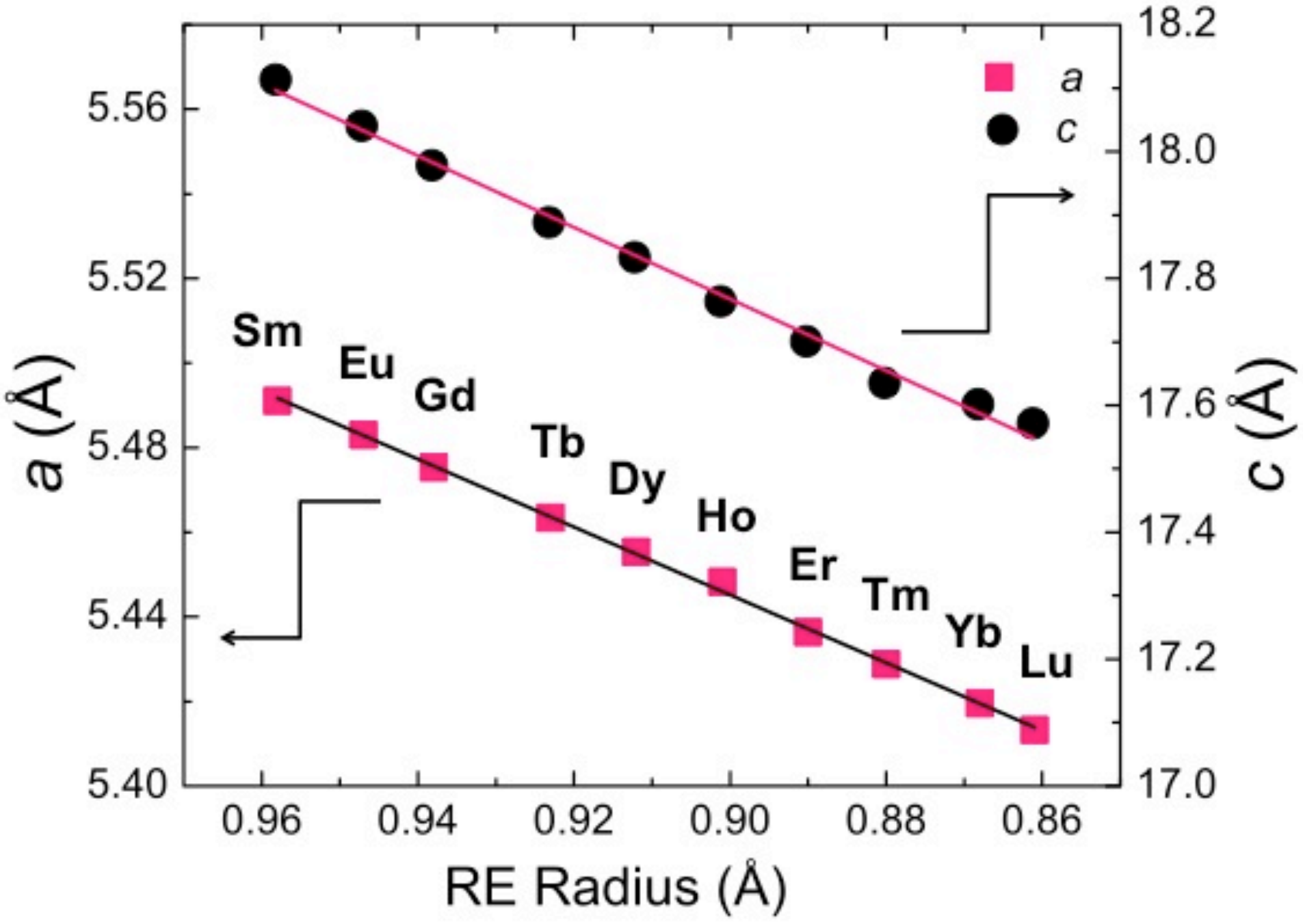

**Figure 6**

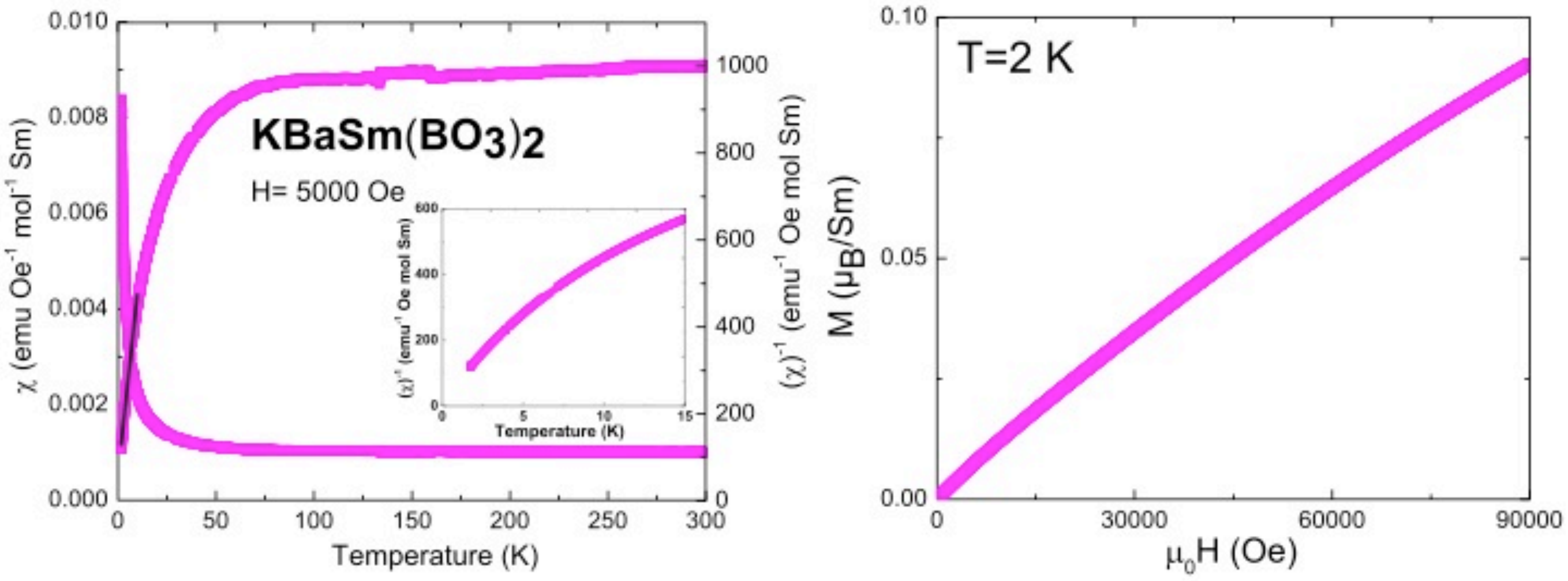

**Figure 7**

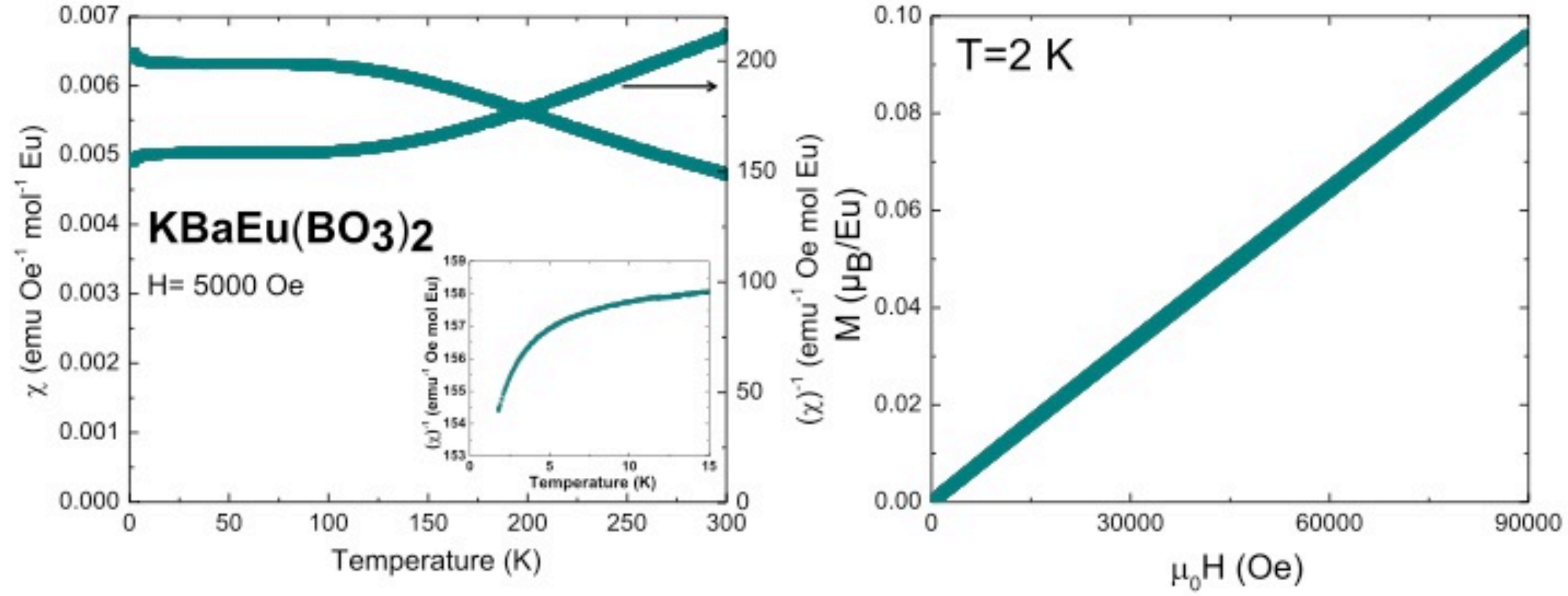

**Figure 8**

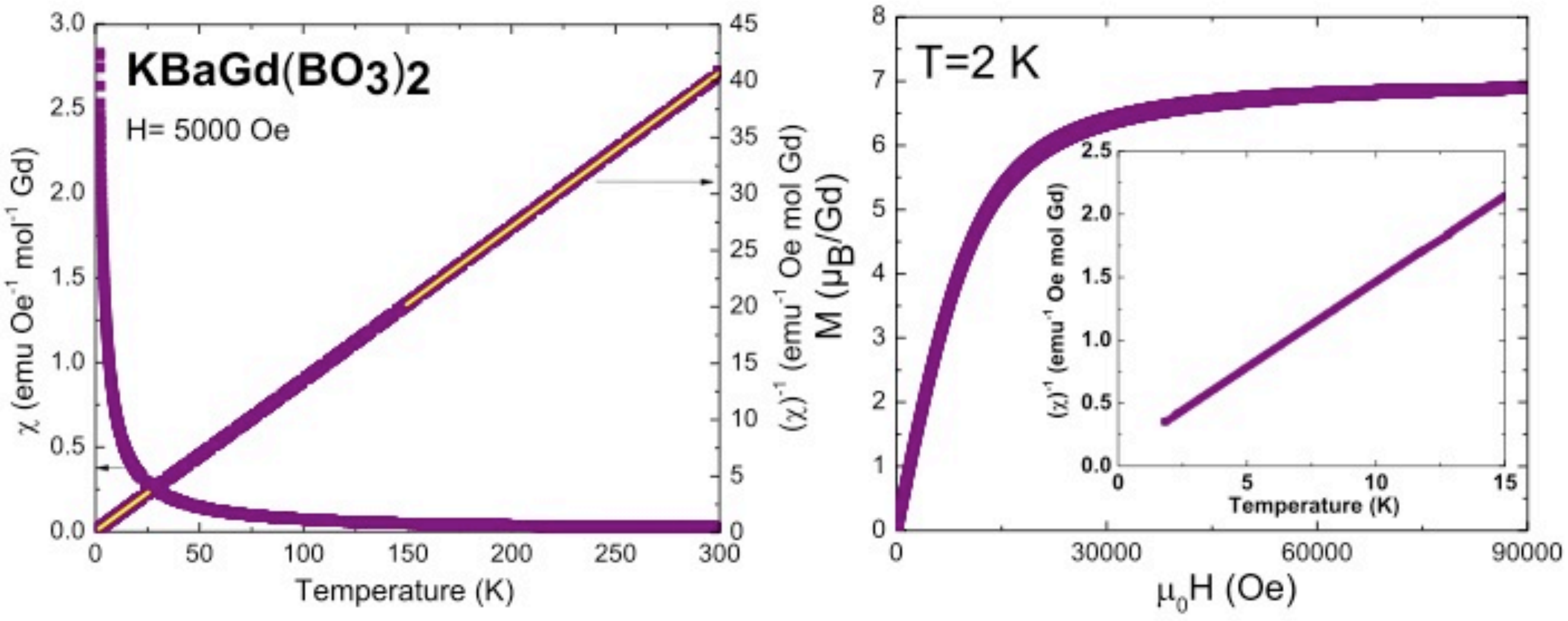

**Figure 9**

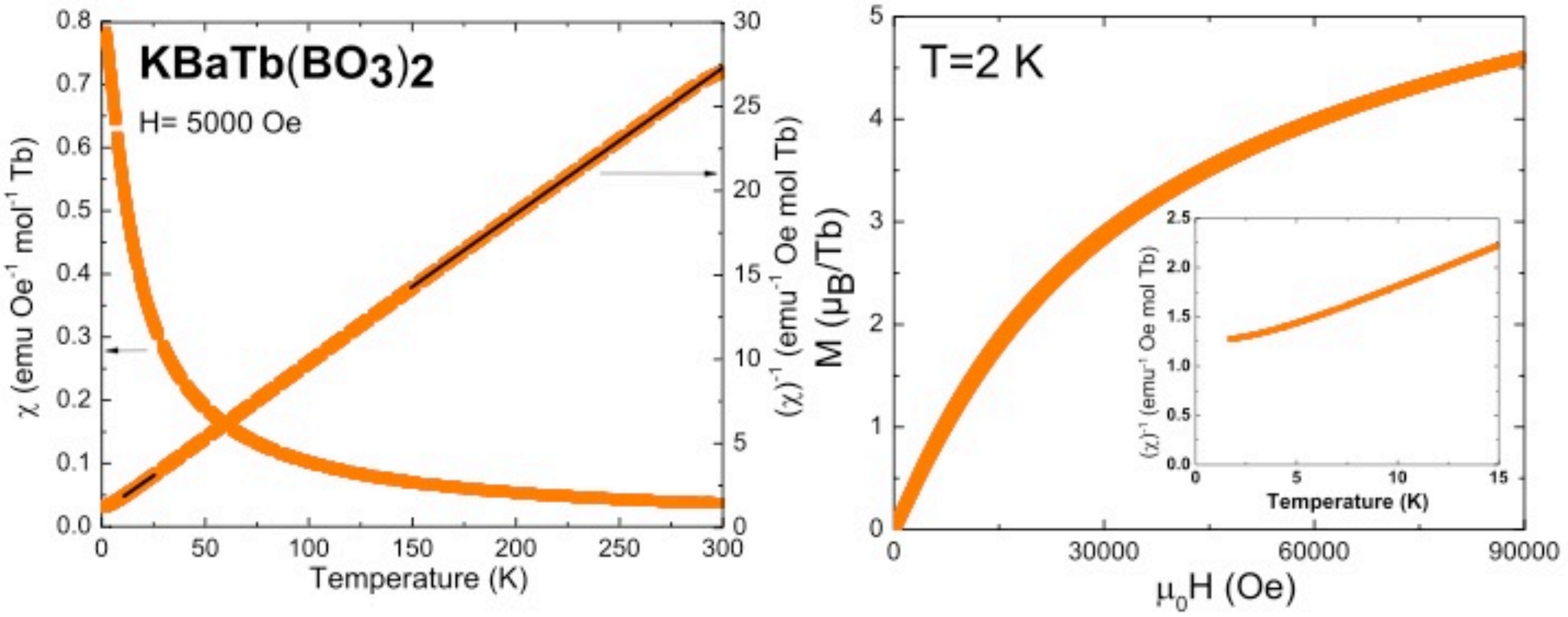

**Figure 10**

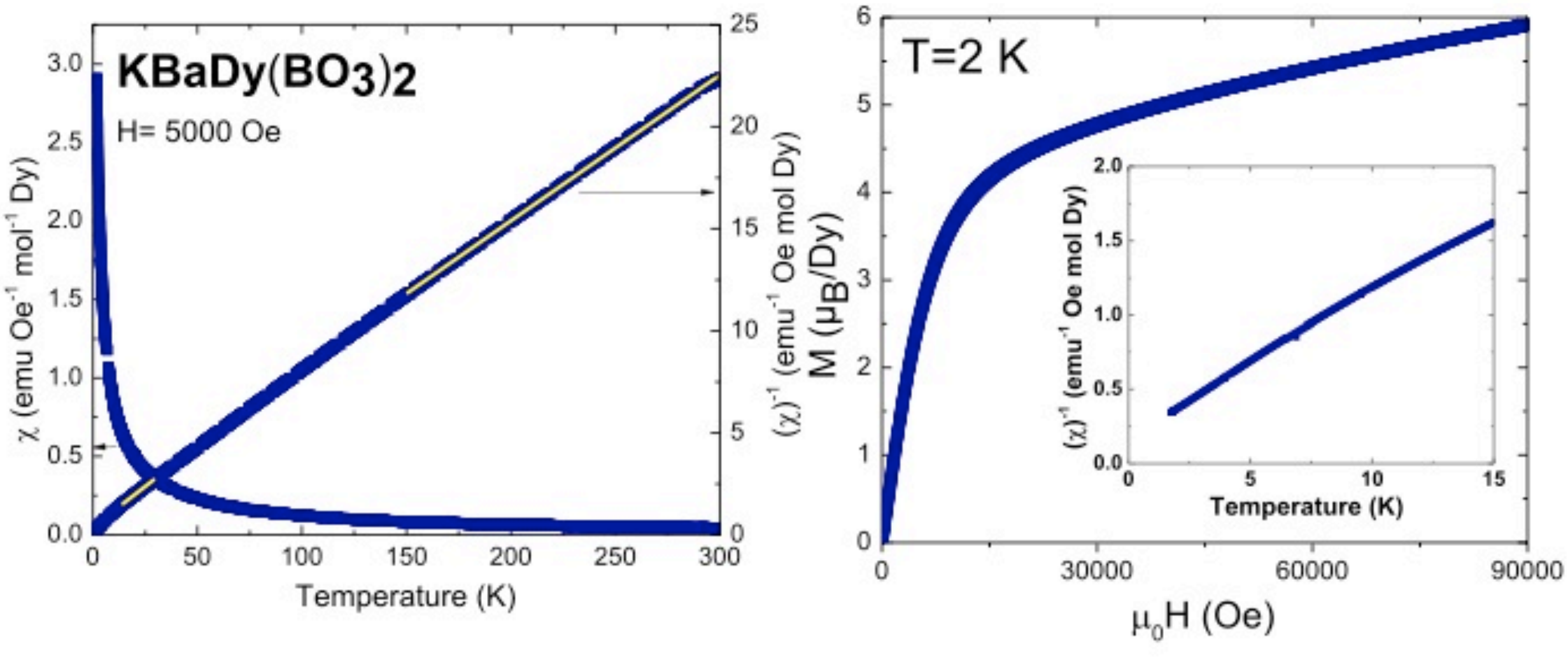

**Figure 11**

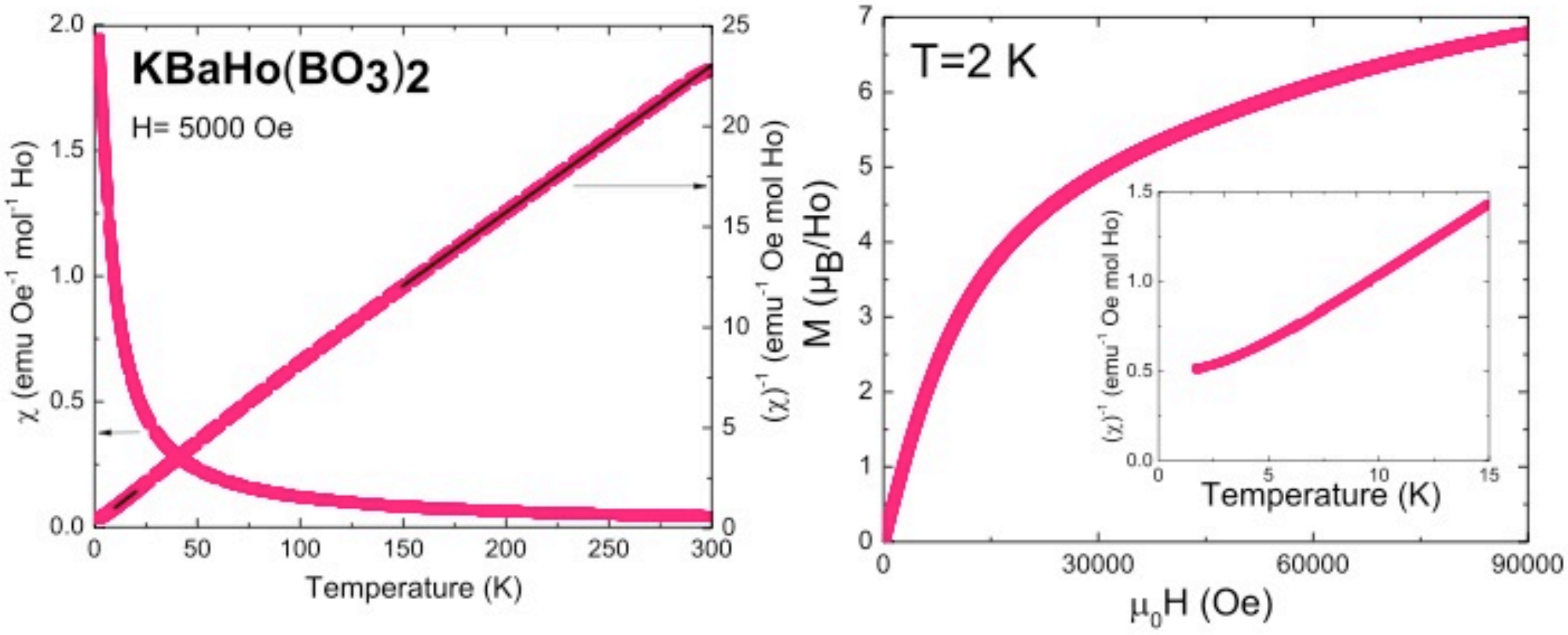

**Figure 12**

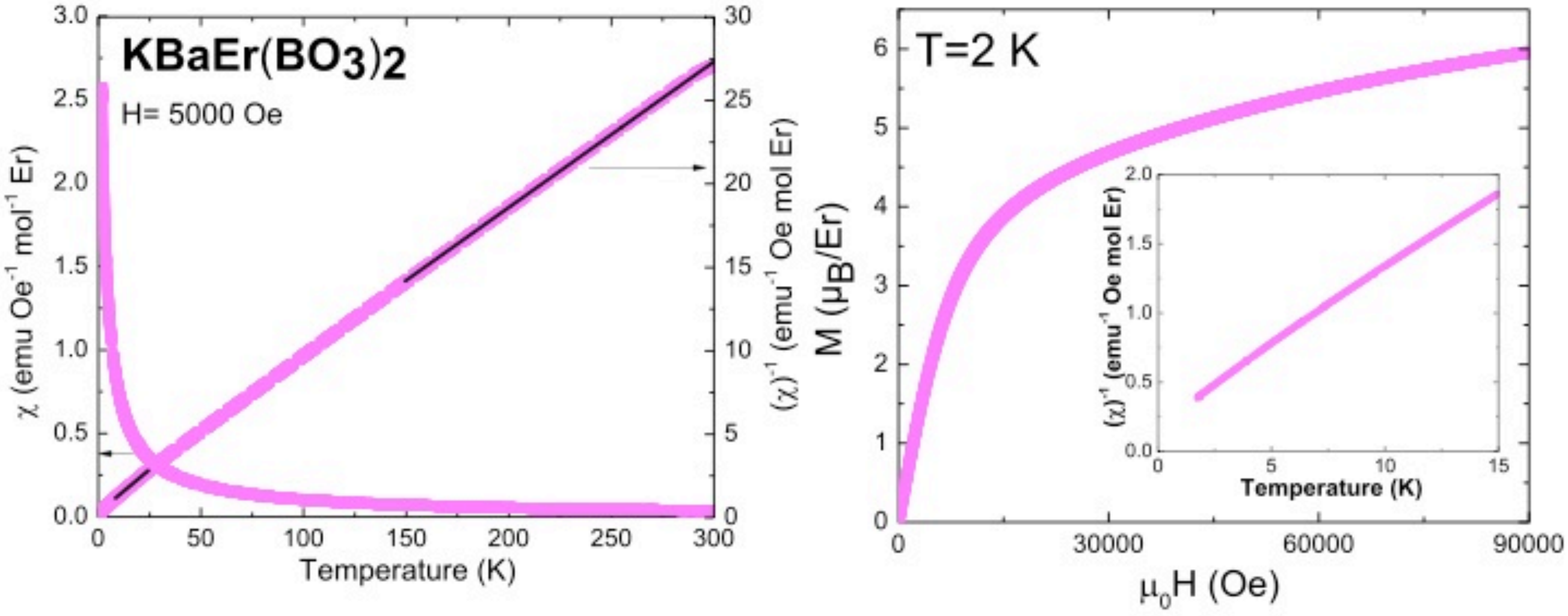

**Figure 13**

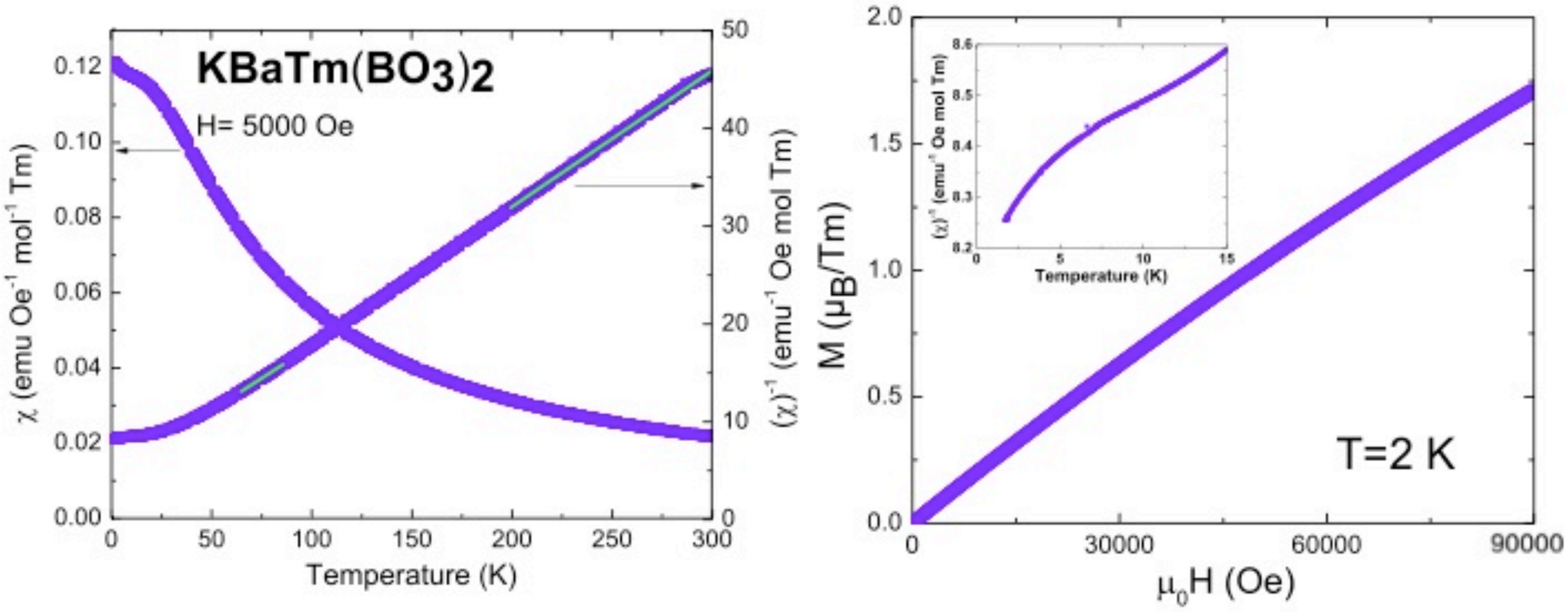

**Figure 14**

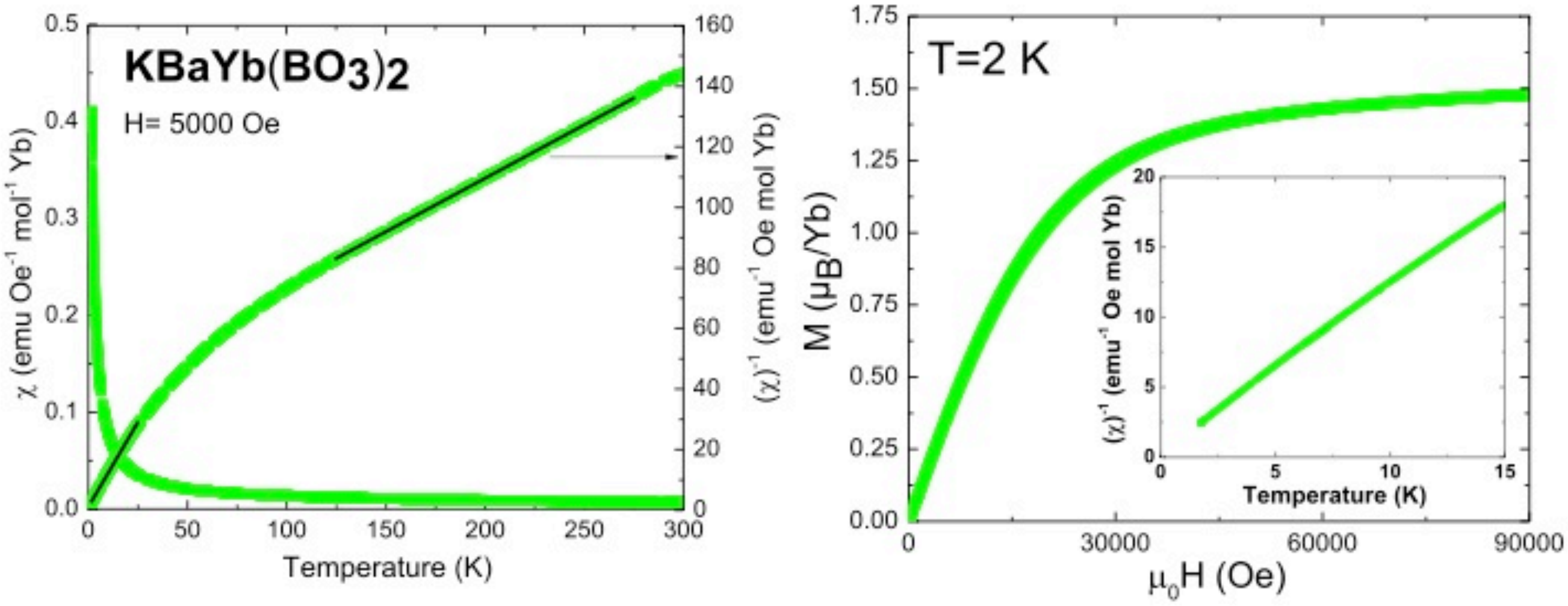

KBaYb(BO3)2
H= 5000 Oe
χ (emu Oe-1 mol-1 Yb)
(χ)-1 (emu-1 Oe mol Yb)
Temperature (K)
T=2 K
M (μB/Yb)
μ0H (Oe)
(χ)-1 (emu-1 Oe mol Yb)
Temperature (K)